# Pulse-Width Modulation Technique With Harmonic Injection in the Modulating Wave and Discontinuous Frequency Modulation for the Carrier Wave for Multilevel Inverters: An Application to the Reduction of Acoustic Noise in Induction Motors

ANTONIO RUIZ-GONZALEZ[1], JUAN-RAMON HEREDIA-LARRUBIA,
MARIO J. MECO-GUTIERREZ, AND FRANCISCO-M. PEREZ-HIDALGO[1]
[1]Department of Electrical Engineering, University of Malaga, 29071 Málaga, Spain
[2]Department of Electronic Technology, University of Malaga, 29071 Málaga, Spain

Corresponding author: Juan-Ramon Heredia-Larrubia (jrheredia@uma.es)

**ABSTRACT** An implementation of a harmonic injection pulse width modulation frequency-modulated triangular carrier (HIPWM-FMTC) control strategy applied to a multilevel power inverter feeding an asynchronous motor is presented. The aim was to justify the reduction in acoustic noise emitted by the machine compared with other strategies in the technical literature. In addition, we checked how the THD at the inverter output was reduced compared to the other control techniques used as a reference. The proposed strategy is based on frequency modulation of the triangular carrier. The main advantage of the proposed method is that only one control parameter is required for modifying the electrical spectrum. Therefore, the mechanical natural frequencies and spatial harmonics of the machine can be avoided, and acoustic noise can be reduced. The effectiveness of the technique was demonstrated after laboratory validation by comparing the acoustic results for a 1 kW motor. The results obtained from the laboratory tests are presented and compared with those of other acoustic measurements using different PWM strategies.

**NOMENCLATURE**

| | |
|---|---|
| $A_M$ | Parameter to adjust the maximum frequency of the carrier. wave |
| f | Fundamental frequency of the modulated wave (50 Hz). |
| $f_k$ | Frequency of the electrical harmonic. |
| f | Frequency of the modulator wave. |
| $f_r$ | Frequency of the radial magnetic force. |
| $f_{sw}$ | Switching frequency. |
| $f_t$ | Frequency of the carrier wave. |
| k | $0, 1, 2, 3\ldots$ |
| K | Truncated level of the function that controls the carrier wave frequency. |
| M | Number of pulses per period. |
| $M_a$ | Amplitude modulation order. |
| $\overline{M}$ | M(t) average value (Frequency modulation order). |
| $m_1$ | Number of stator phases. |
| p | Number of pairs of poles. |
| r | Vibration mode order. |
| s | Slip. |

| s₁ | Stator slots. |
| s₂ | Rotor slots. |
| THD | Total harmonic distortion. |

## I. INTRODUCTION

Multilevel Power Inverters (MLI) refer to the connection of individual inverters called stages to provide an output voltage as a function of the association between the voltages of different stages. Increasing the number of levels significantly reduces harmonic distortion, both in the voltage and current [1]. The multilevel inverter provides a suitable solution for medium- and high-power systems to synthesize an output voltage that makes possible a reduction of voltage and current harmonic content. Multilevel converters have superior characteristics compared with two-level converters, such as lower dv/dt, lower switching losses, and better output waveform quality. With the ability to handle high-voltage and high-power devices, multilevel converters are widely used in medium- and high-voltage (>3 kV) power conversion systems but are also suitable for low-voltage power applications (approximately 400 V) owing to the reduced output filter volume or the higher fault tolerance capability they can achieve [2].

Although there is a wide diversity of multilevel inverter topologies [3], [4], the most common voltage-source topologies are neutral-point diode converters (NPC) [5], [6], [7], [8], flying capacitor converters (FCC) [9], cascaded H-bridge converters (CHB) [10], [11], and modular multilevel converters (MMC) [12], [13], [14].

A multilevel cascaded H-bridge inverter has several advantages over other topologies (e.g., flying capacitor or clamped capacitor, diode-clamped, and neutral-point clamped). The CHB inverter is modular and simple to control, because each cell has the same structure. In addition, the CHB inverter makes it possible to reach many levels with minimum necessary components. These components mitigate the switching losses of the devices used and increase the reliability and efficiency of the circuit [15]. Therefore, this topology is widely used in industrial applications [4]. These cascaded multilevel inverters are widely used in renewable energy systems owing to several advantages, such as the absence of voltage imbalance problems, possible elimination of the DC-DC converter, low-frequency switching enabling cheaper transistors, and the absence of floating capacitors and latching diodes. The disadvantage is that powering cascaded multilevel inverters requires separate power supplies for the different stages of the multilevel inverters.

Numerous modulation techniques for the control of MLIs are available in the literature, such as carrier-based pulse-width modulation (PWM) [16], space vectors [17], [18], and selective harmonic elimination [19], [20]. Carrier-based modulation is a popular technique for networked MLIs such as carrier phase-shifted pulse-width modulation (PS-PWM) and carrier-level-shifted PWM (LS-PWM) [21], [22]. Other techniques modify the frequency of the carrier signals and amplitude of the harmonics of the modulating signal to reduce losses and improve the electrical parameters of the modulated wave [23].

In [24], the reference signal was shifted vertically to obtain a switching signal for MLI. Because there is a level shift between the carrier signals in LS-PWM, there is a higher voltage distortion, which is not the case in PS-PWM. The proposed strategy focuses on cascaded or H-bridge inverters and follows an efficient PS-PWM technique, which not only imposes a uniform power distribution between the cells, but also leads to lower voltage distortion and effective current harmonic distortion. However, it is well known that the reduction in noise emitted by a motor depends on the modulation technique [25], [26], [27]. In the technical literature, to the best of our knowledge, there is no recent work on induction motor noise reduction using multilevel inverters.

An application of this type of CHB inverter is the power supply of high-power motors fed at high voltages to reduce the current in the windings, which require several stages for each phase. The increase in noise emitted by AC electrical machines when powered by such devices to control speed or torque is well known.

Fig. 1.a) shows the generic topology of a three-phase 5-level multilevel. Fig. 1. b) shows the details of the H-bridge with its four control signals: IGBT11, IGBT13, IGBT15, and IGBT17. It should be noted that the trigger signals corresponding to the other two bridges, H2 and H3, are 120° and 240° out of phase, respectively.

All electric drives increase the acoustic noise of the motors. Depending on the control technique used, the level of acoustic noise can be reduced [28]. In this study, a control technique that significantly attenuates the acoustic noise produced by a motor was proposed. The acoustic results and inverter output wave quality parameters of the proposed modulation technique were compared with those of other techniques in scientific literature, which showed good results. Specifically, these are amplitude-shifted modulation in the carrier waves (Fig. 2a), phase-shifted modulation in the carrier waves (Fig. 2b), and phase-shifted modulation with an injection of harmonics in the modulating wave (Fig. 2c).

Chapter II summarizes the frequencies of the acoustic noise that can be identified after feeding an induction motor with either a sine wave supply or an inverter that generates a set of identifiable harmonics.

Chapter III discusses how a strategy is defined when applied to multilevel inverters, by showing its basic concepts. Subsequently, the genesis of the signals attacking the gates of the IGBTs and the waveform at the inverter output are presented as functions of control parameters. Finally, the evolution of the instantaneous frequency of the carrier wave with respect to the modulating signal was presented as a function of the control parameter.

Section IV presents the laboratory results. It shows the equipment necessary to generate the waves that control the multilevel inverter for the different strategies tested, the layout of the sound-level meter, and the signal analyzer that records the waveforms of the voltages and currents and their corresponding electrical spectra. Finally, Section V presents the conclusions.

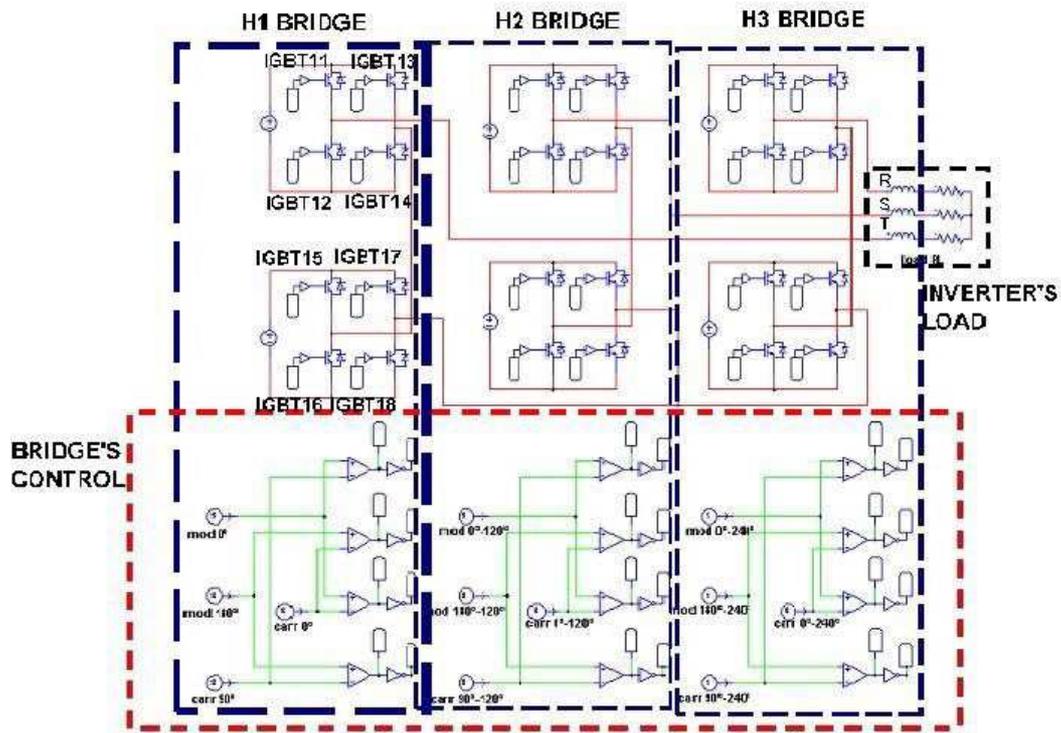

(a)

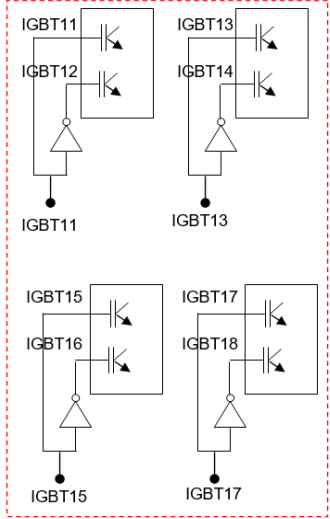

(b)

**FIGURE 1.** a) Multilevel inverter H–bridge. b) Control signals for a H–bridge.

## II. ACOUSTIC NOISE IN INDUCTION MACHINES

Motor acoustic noise can be classified into two types depending on its origin. The first type comprises aerodynamic and mechanical noise (fan, bearing, and magnetostrictive expansion of iron). They are low-frequency, depend on the degree

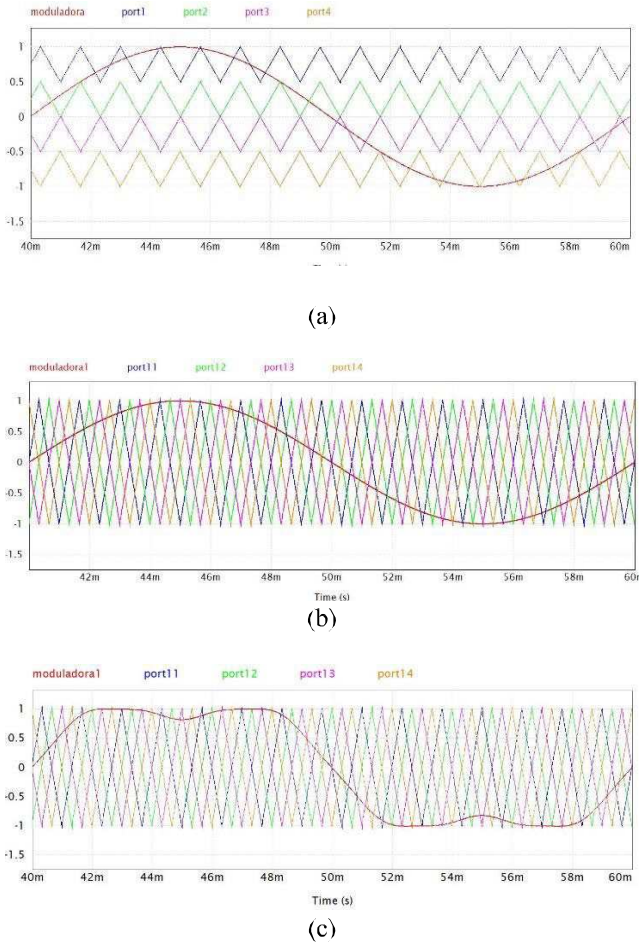

**FIGURE 2.** Multilevel inverter modulation: a) with amplitude shift; b) with phase shift; c) with phase shift and injection of harmonics into the modulating signal.

of saturation of the magnetic circuit, and do not depend on load level.

The other type (spatial and temporal harmonics) is electromagnetic noise, which is a consequence of the finite number of slots and coils of the electrical machine (spatial harmonics) and the harmonics of the output voltage of the inverters that feed it (temporal harmonics). The latter depends on the load level and degree of saturation of the magnetic circuit.

If the motor is fed from a balanced three-phase power supply, the only harmonics that appear are spatial harmonics. These are proportional to $B^2(\alpha, t)$, where B is the magnetic flux density in the air gap, $\alpha$ is the angular position of the reference, and t is time. Other harmonics are caused by magnetostriction, bearing noise, and aerodynamics, which depend on the speed and the number of fan blades on the shaft.

If the motor is fed from an inverter, the radial forces that appear are a consequence of the products of stator harmonics with the same number n as their time harmonic, reaching a frequency of $f_r = 2f \cdot (2k \cdot m_1 \pm 1)$. However, the frequencies of the harmonics of the rotor currents of the same number n of its time harmonic are n times $2f \cdot (s_2/p) \pm 1)$, where $n = 2 \cdot k \cdot m_1 \pm 1$ [29].

However, the most important acoustic harmonics that appear are produced by the interaction of the switching frequency of the carrier signal, $f_{sw}$, with the first-order time harmonics. These acoustic harmonics have a vibrational mode (r = 0) of high acoustic response at a vibrational frequency $f_r = | \pm (a \cdot f_{sw} \pm b \cdot f) - f |$, where f is the fundamental component of the frequency of the modulated waveform and $f_r$ is the acoustic harmonic. If a is even, b is odd, and vice versa; that is, $f_{sw} \pm 2f$, $f_{sw} \pm 4f$, $2f_{sw} \pm f$, $2f_{sw} \pm 3f \ldots$ [29]. In other words, the acoustic harmonics produced by the inverter are expected to appear at a frequency of one plus or minus the frequency of the electrical spectrum of the modulated wave applied to the motor. If there is an electrical harmonic at a frequency of $n \cdot f$, where n is a natural number, then the temporal type of acoustic harmonic appears at a frequency of $(n \cdot f + 1)$ or $(n \cdot f - 1)$.

Power inverters regulate the voltage level of the machine, as well as the frequency, when it is intended to control the speed using a specific control strategy, generating specific electrical harmonics whose amplitude and location in the spectrum depend on this strategy. The existence of these harmonics produces extra heating, which results in a reduction in performance, as well as pulsating torques that cause acoustic noise and vibrations in the structure of the machine.

To reduce the effect of the electrical harmonics on the motor, the frequency of the carrier signal was increased by limiting it to the maximum switching frequency. For large powers and voltages, owing to the reduced cost of the switches, several elementary stages can be connected in series, requiring one or several different sources. This reduces the number of commutations for each cycle of each switch, and can further increase the voltage and power of the electrical machine.

To increase the voltage of the fundamental term, a sine wave signal with harmonic injection is used as the modulating signal [30], [31], [32]. The vibration modes when the motor is supplied with a balanced three-phase wave are variable, and the acoustic level emitted by the motor with these supplies is limited by the international standards. However, when a power inverter is installed and the load level is high, time harmonics considerably affect the overall acoustic level produced when the vibration mode is low. In such cases, the control strategy is particularly relevant.

### III. NEW HIPWM-FMTC STRATEGY APPLIED TO MULTILEVEL INVERTERS

The HIPWM-FMTC technique compares a harmonic-injection modulating wave with a frequency-modulated triangular carrier wave to generate an output waveform. The instantaneous frequency of the carrier wave was adjusted according to a periodic function synchronized with the modulating wave. The main motivation for using this technique compared to the classic PWM sinusoidal technique is the reduction of the total harmonic distortion, the reduction of the distortion factor, and the shift of temporal harmonics to higher frequencies for any modulation frequency order, avoiding their coincidence in frequency with the mechanical resonances of the motor structure.

For one period of the modulating waveform, the number of switching pulses is increased at an interval around the highest slope of the modulating waveform and reduced to zero when the modulating waveform has its lowest slope (around its maxima and minima), maintaining a constant number of switching pulses for a conventional PWM technique. The maximum switching speed is determined based on the theoretical limit of the inverter. This technique has been referred to by the technical literature for 3-level inverters as HIPWM-FMTC (Harmonic Injection PWM- Frequency modulated Triangular Carrier) [28], [30], [31], [33], [34]. A modification of the HIPWM-FMTC technique for three levels was proposed by introducing a parameter to truncate the triangular function to increase the instantaneous switching frequency and shift the electrical harmonics to higher frequencies [32]. The technique published in [32], but applied to multilevel inverters, is called truncated HIPWM-FMTC or HIPWM-FMTCt. Thus, the instantaneous pulsation of the carrier signal $\omega_i$ is a synchronized discontinuous function of the fundamental term of the modulating wave, defined as follows:

$$\omega = \frac{d\theta}{dt} = A_M \cdot \omega_m [cos^2 \omega_m t - K] \quad (1)$$

In the intervals in which (1) takes negative values, $\omega_i$ is forced to zero (switching is blocked). The desired modulating wave and the instantaneous pulsation of the carrier wave $\omega_i$ must be synchronized, as shown in Fig. 3. Parameter K (truncation level) is defined as a real number in the interval [0, 1], and it does not depend on the value of $\overline{M}$ (number of pulses per period). $A_M$ is a parameter that controls the maximum frequency of a carrier wave. These parameters allow the electrical spectrum to be modified without changing the value of $M$ during the modulated wave period. After setting the value of $\overline{M}$, for each value of K, it is necessary to obtain $A_M$.

The relative frequency of the carrier and modulating waves is the instantaneous frequency modulation order (null between $t_1$ and $t_2$ and between $t_3$ and $t_4$ for each period of the modulating wave):

$$M(t) = \frac{\omega_i}{\omega_m} \quad (2)$$

The average value of M(t), $\overline{M}$, defines the frequency modulation order. For a complete period of the modulating wave, both the M and $\overline{M}$ values are the same:

$$\overline{M} = \frac{1}{T_m} \int_0^{T_m} A_M [cos^2(\omega_m t) - K] dt \quad (3)$$

with $T_m = 2\pi/\omega_m$. The first zero of (1) is reached for $t = t_1$ with $t_1 = (cos^{-1}\sqrt{K})/\omega_m$

To achieve synchronization between the modulator and carrier waveforms, $M$ must be set to an integer value. To avoid even and triple harmonics with three-phase inverters, must be set to odd multiples of three K y $A_M$ from equation (3), which makes synchronization between the modulating and carrier waveforms possible. The objective of this synchronization is to ensure that the maximum instantaneous value of the carrier frequency coincides with the instant at which the modulating sine wave with harmonic injection exhibits a higher slope (angles close to 0 and π rad). Its value is then reduced following a quadratic cosine function around these phases, and reaches zero during a time interval that depends on K, as shown in (1) and (3).

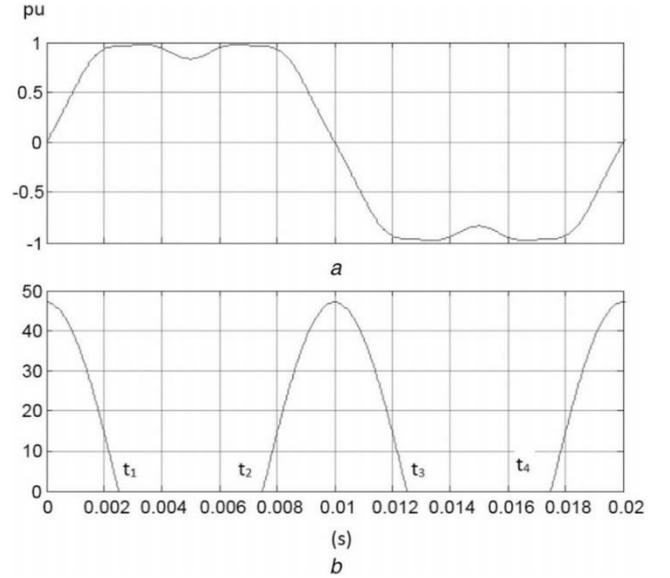

**FIGURE 3.** Temporal synchronization between modulating wave and the instantaneous modulation order for the frequency of the carrier wave: a) Modulating wave versus time; b) Instantaneous modulation order versus time (fm = 50 Hz, K = 0.5 and $A_M$ = 30π, $M$ = 15).

For K = 0, the carrier wave frequency is theoretically zero only for instants corresponding to the phase values of the modulating waves π/2 and 3π/2 (0.005 and 0.015 s, respectively, for 50 Hz). The same carrier wave can be obtained using the HIPWM-FMTC technique [30], where $\omega_c$ is $30\omega_m$ and $k_f$ is 0, and $\omega_c$ and $k_f$ are the control parameters of the proposed technique. The waveform obtained at the inverter output was the same for both techniques for both control parameter values.

For any other value of K ϵ R/ (0, 1],
of the proposed technique is zero at time intervals near the π/2 and 3π/2 phase values of the modulating wave. The instantaneous frequency of the carrier wave increased around the 0 and π rad phases of the modulating wave, coinciding with the steepest slope. The modulating waveform changes its slope more rapidly for these phase values. Therefore, increasing the number of pulses around the phases is desirable to transmit more information to the modulated waveform.

As shown in Fig. 4, K = 0.5 (the center value of K), and $M$ = 15 ($A_M$ = 30π), During half of this period, there was a modulated triangular carrier and during the other half, switching was prevented. Therefore, the inverter switches during only half of the total period of the modulating waveform, that is, 20 ms (50 Hz). The maximum carrier frequency excursion is $A_M \omega_m \cdot (1 - K) = 15\pi \cdot \omega_m$, as shown in Eq.(1)–(3). Therefore, $t_1$ was 2.5 ms, and $t_2$, $t_3$, and $t_4$ were 7.5, 12.5 y 17.5 ms, respectively. The pulses are concentrated around instant mT/2, where T is the fundamental period of the modulating wave and m is the series of natural numbers. At these times, the instantaneous frequency of the carrier wave has a maximum value of $30\pi/2$ times the frequency of the modulating wave instead of 15, as would be the case in the Sinusoidal PWM technique, and the area under the instantaneous modulation order curve in Fig. 3. b) is $\overline{M} = 15$. Each value of K creates a different $A_M$ value when is set.

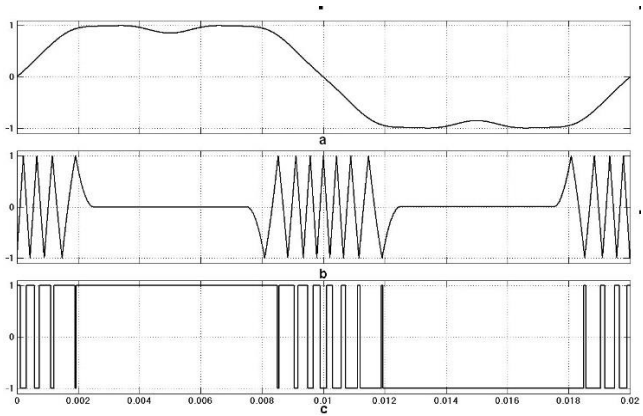

**FIGURE 4.** Simulation of the waves used in the generation of the proposed PWM. a) Modulating wave with harmonic injection for f = 50 Hz; b) Carrier signal HIPWM–FMTCt technique (K = 0.5 and $A_M = 30\pi$); c) Line–to–neutral voltage at the output inverter for a single bridge H.

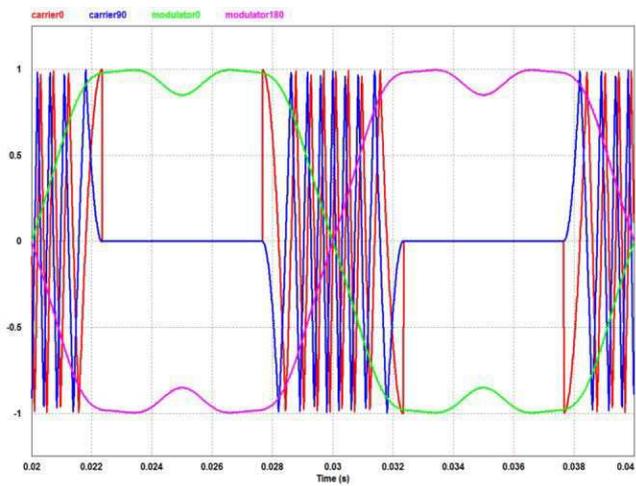

**FIGURE 5.** Simulation of the waves used in the generation of the proposed PWM: Modulating and carrier waves with harmonic injection for f = 50 Hz for HIPWM–FMTCt technique (K = 0.5 and $A_M = 30\pi$).

The modulation technique is shown in Fig. 5 and 6 with reference to the power structure shown in Fig. 1. Fig. 7 shows the different signals driving the gates of the 2-stage H-bridge. The pulse train that excites the base of IGBT11 is obtained by comparing the modulating signal with the phase 0° harmonic injection with the phase 0° carrier. In turn, the pulse train that excites the base of IGBT13 is obtained by comparing the modulating signal with a 180° phase harmonic injection with the 0° phase carrier. IGBT15 was excited with pulses resulting from a comparison between the modulator and carrier phases at 0° and 90°, respectively. Compared with the carrier at 90°, the 180° phase modulator will in turn provides pulses that drive the gate of the IGBT17.

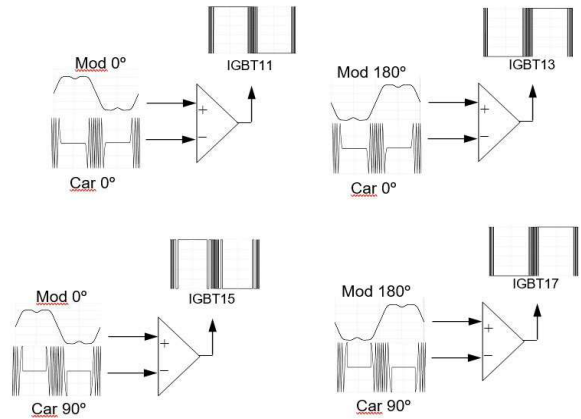

**FIGURE 6.** Simulation of the generation of the trigger pulses for an H–bridge IGBT.

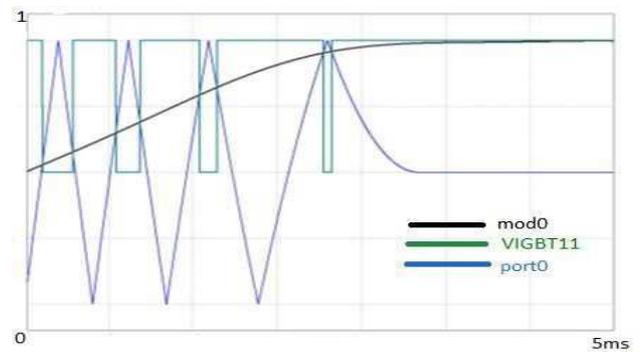

(a)

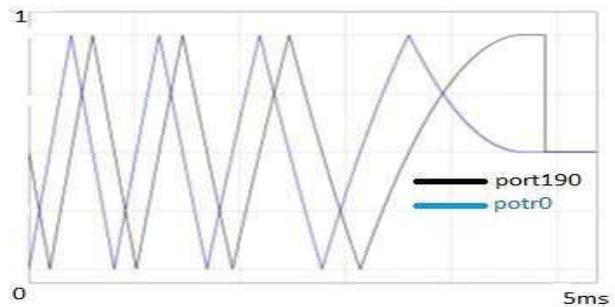

(b)

**FIGURE 7.** a) Simulation of the generation of the trigger pulses for an H–bridge IGBT; b) Simulation of the generation of two carrier waves shifted by 90°.

Table 1 shows the evolution of some variables against K when $\omega_m$ and $\overline{M}$ were set to 100π rad/s and 15, respectively. The variables presented are $t_1$, which represents a quarter of the time that enables per-period switching of the output waveform (see Fig. 3); $A_M$, to set the maximum frequency of the carrier waveform; and the maximum frequency modulation order, which is calculated as $A_M \cdot (1 - K)$. The maximum carrier frequency is $A_M \cdot (1 - K) \cdot \omega_m$.

**TABLE 1.** $A_M$, $T_1$ and $A_M \cdot (1-K)$ for different values of K.

| K | $A_M$ | $t_1$ (ms) | $A_M (1 - K)$ |
|---|---|---|---|
| 0.2 | 44.277 | 3.524 | 35.422 |
| 0.3 | 55.134 | 3.155 | 38.594 |
| 0.4 | 70.638 | 2.820 | 42.383 |
| 0.5 | 30π | 2.5 | 47.124 |
| 0.6 | 133.513 | 2.180 | 53.405 |
| 0.7 | 208.142 | 1.845 | 62.443 |
| 0.8 | 386.859 | 1.476 | 77.372 |

Fig. 8 shows the phase-to-phase voltage of a single phase with nine voltage steps with the proposed new technique and for a value of K = 0.5 and f = 50 Hz.

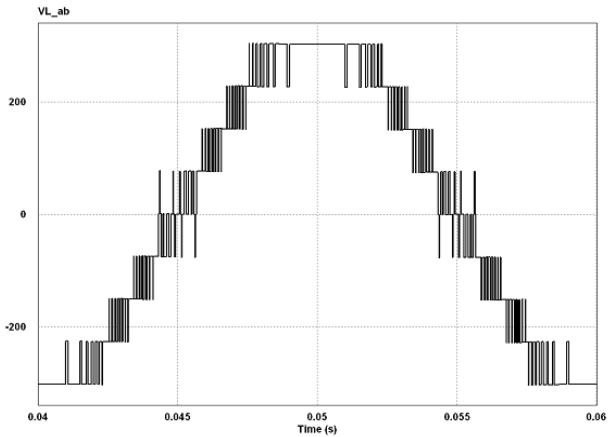

**FIGURE 8.** Phase-to-neutral modulated voltage waveform of a multilevel inverter H-bridge with HIPWM-FMTCt strategy (K = 0.5, f = 50 Hz and $M$ = 15).

## IV. RESULTS

The asynchronous motor used was an AEG™ 220/380 V, 1 kW, 4-pole (p = 2) motor with 36 slots in the stator ($s_1$) and 26 slots in the rotor ($s_2$). The sound emitted by the motor was measured in a semi-anechoic chamber at a distance of 1 m between the microphone and machine in the transverse direction of the longitudinal axis. For the multi-level inverter implementation, the GPT-IGBT module from GUASCH S.A™ was used. It offers a compact and versatile insulated-gate bipolar transistor (IGBT) power stack for motor control. This power module includes a 3-phase rectifier bridge, capacitor bank, IGBTs with a forced-air-cooled heat sink, optocoupled drivers, output phase current sensors, DC-Link current sensor, and DC-Link voltage sensor. The main electrical features are as follows: maximum voltage applied to the DC link, 750 V, and output current per phase, 32 A.

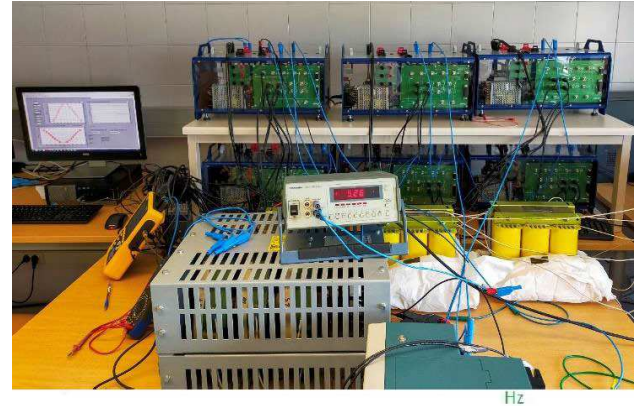

(a)

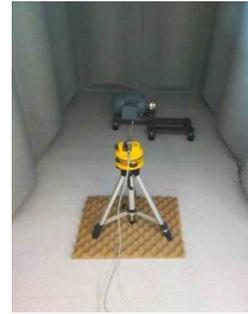 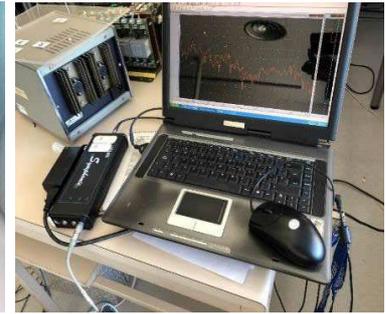

(b)　　　　　　　(c)

**FIGURE 9.** a) Multilevel inverter and network analyzer, NI9154 control equipment, b) Semi-anechoic chamber, c) 01-dB METRAVIT Symphonie sound level meter.

To generate the control signals for the inverter stages, hardware from National Instruments, specifically the NI9154 card, was used, and a Labview™-based platform was developed for the generation of different PWM techniques. The current harmonics at the inverter output were measured using three-phase power analyzers, and the energy quality was measured using a Chauvin Arnoux™ C.A 8336, that measures the first 50 harmonics. The acoustic noise emitted by the engine was measured and analyzed using a 01-dB METRAVIT Symphonie model sound-level meter with a G.R.A.S.™ 40FA microphone of 50 mV/Pa at 250 Hz, which meets the requirements of the IEC 1094 standard, together with a G.R.A.S.™ type 26AK preamplifier.

Fig. 9 a) b) and c) illustrates the multilevel inverter, the network analyzer, the NI9154 control equipment, and the semi-anechoic chamber.

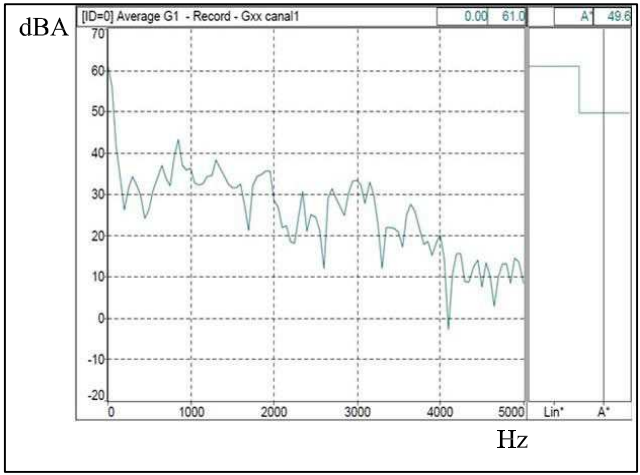

**FIGURE 10.** Noise level (dBA)/Hz of the motor powered by a three-phase balanced 220 V grid.

**TABLE 2.** Number of the spatial harmonics and frequency for 50 Hz modulating wave.

| k | Type 1 | Type 1 | Type 3 | Type 3 | Type 3 |
|---|--------|--------|--------|--------|--------|
| 1 | 28f<br>1200 Hz | 24f<br>1400 Hz | 11f<br>550 Hz | 13f<br>650 Hz | 15f<br>750 Hz |
| 2 | 54f<br>2700 Hz | 50f<br>2500 Hz | 24f<br>1200 Hz | 26f<br>1300 Hz | 28f<br>1400 Hz |
| 3 | 80f<br>4000 Hz | 76f<br>3800 Hz | 37f<br>1850 Hz | 39f<br>1950 Hz | 41f<br>2050 Hz |
| 4 | 106f<br>5300 Hz | 102f<br>5100 Hz | 50f<br>2500 Hz | 52f<br>2600 Hz | 54f<br>2700 Hz |
| 5 | 132f<br>6600 Hz | 128f<br>6400 Hz | 63f<br>3150 Hz | 65f<br>3250 Hz | 67f<br>3350 Hz |

The following figures correspond to the acoustic results of the motor with a sinusoidal power supply, waveforms of the line voltages with the modulation techniques SPWM1, SPWM2, and SPWM3, and frequency spectra of the motor line current for each of them. In addition, the line voltage obtained using the proposed modulation technique is shown for a value of K that minimizes acoustic noise emitted by the motor. Fig. 10 shows the noise spectrum emitted by the motor when it was fed by a three-phase sinusoidal power supply of 220 V and 50 Hz.

In Table 2 the different space harmonics can be identified:

1) Acoustic noise due to the products of the rotor space harmonics of the same harmonic number, with frequencies $f_r = 2f \cdot [1 \pm k \cdot (s_2/p)(1-s)]$, with $k = 0, 1, 2, 3, \ldots$

2) Electromagnetic noise owing to the products of the stator space harmonics of the same harmonic number, whose frequency of the radial magnetic forces is 2f (100 Hz) and $r = 4$.

3) Interaction between stator and rotor space harmonics: $f_r = f \cdot [k \cdot (s_2/p)(1-s)]$ and $f_r = f \cdot [k((s_2/p)(1-s) \pm 2)]$

Comparative studies will be conducted between various modulation strategies for inverter switches that feed motors.

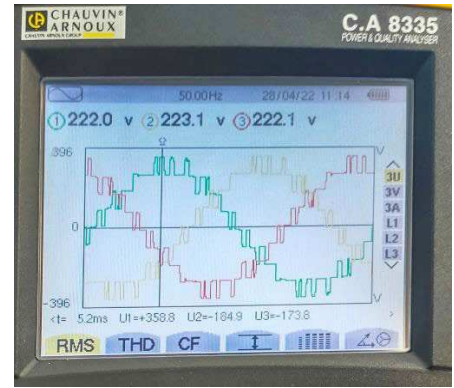

(a)

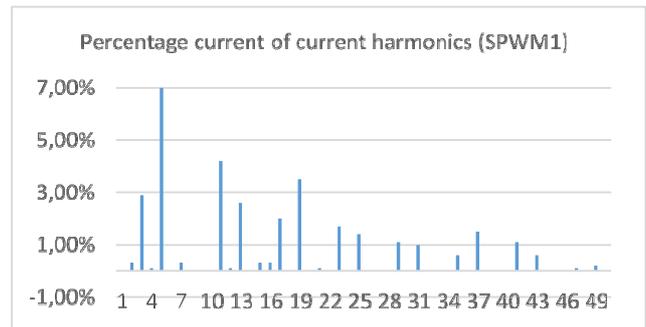

(b)

**FIGURE 11.** a) Modulated waves for the multilevel inverter line signal with SPWM1; b) Percentage electrical spectrum of the output current of the SPWM1 multilevel inverter: V = 220 v RMS of the fundamental term.

Three techniques from technical literature on multilevel converters were applied. SPWM1 is an amplitude-shifted technique for a triangular carrier wave with a sine wave modulator for which the output line voltage and the current frequency spectrum are displayed in the Fig. 11.

The line voltage and current–frequency spectra are presented. Fig. 11 (a) shows the modulated waves for the multilevel inverter line signal with SPWM1; b) Percentage electrical spectrum of the output current of the SPWM1 multilevel inverter: V = 220 V RMS of the fundamental term.

The SPWM2 technique involves a phase shift of triangular wave carriers and sine modulator waves. See Fig. 12: a) Modulated waves for the multilevel inverter line signal with SPWM2; b) Percentage electrical spectrum of the output current of the SPWM2 multilevel inverter: V = 220 v RMS of the fundamental term.

The SPWM3 technique consists of phase-shifting for a triangular wave carrier and injecting harmonics into the modulating wave. Fig. 13 (a) shows the modulated waves for the multilevel inverter line signal with SPWM3 and b) shows the percentage electrical spectrum of the output current of the SPWM3 multilevel inverter: V = 220 V RMS of the fundamental term.

Finally, the proposed HIPWM-FMTCt strategy with the peculiarity of modulating and truncating the carrier wave frequency is given in Fig. 14 HIPWM-FMTCt with K = 0.55 and

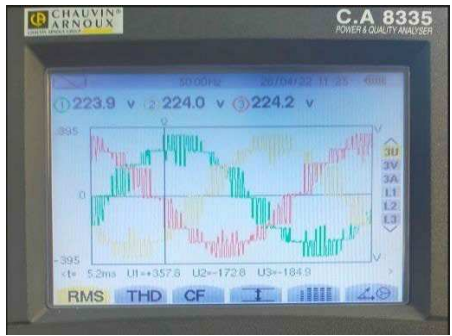

(a)

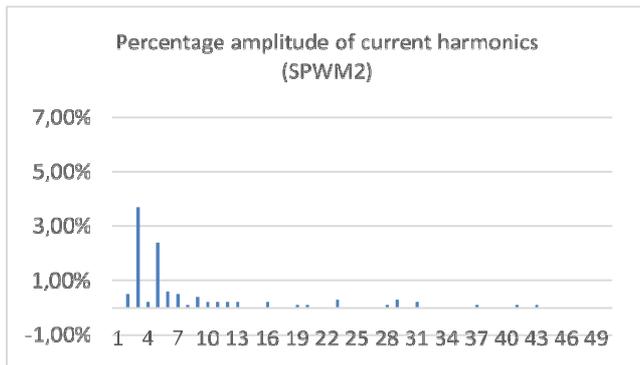

(b)

**FIGURE 12.** a) Modulated waves for the multilevel inverter line signal with SPWM2; b) Percentage electrical spectrum of the output current of the SPWM2 multilevel inverter: V = 220 v RMS of fundamental term.

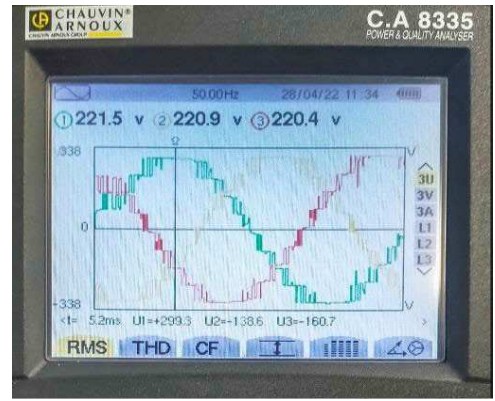

(a)

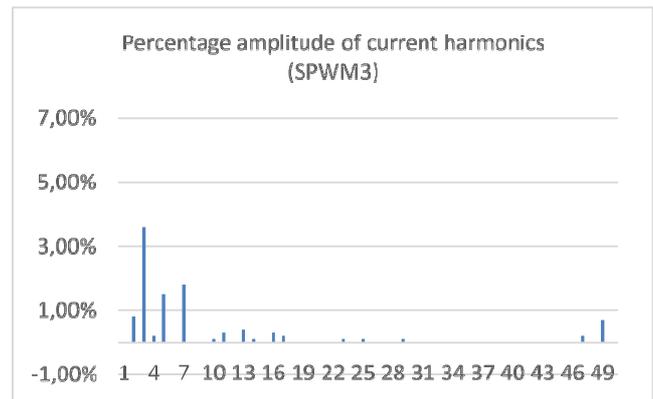

(b)

**FIGURE 13.** a) modulated waves for the multilevel inverter line signal with SPWM3; b) Percentage electrical spectrum of the output current of the SPWM3 multilevel inverter: V = 220 v RMS of fundamental term.

V = 220 v RMS of fundamental term: a) Waveform at the output of the multilevel inverter for the HIPWM-FMTCt; b) Percentage frequency spectrum: current.

Since the line voltage obtained is different for each strategy, from the same DC-link voltage value, the criterion used to establish a comparison between the different control strategies is the RMS voltage of the fundamental term, which is set at 220 V.

Fig. 15 shows the results of the SPWM1 strategy with M = 15. In addition to the noise caused by space harmonics, time harmonics also appeared. It can be observed that electrical harmonics 5, 11 (1·750-4·50 Hz), 19 (1·750+4·50 Hz), 25, and 37 generate sound harmonics 6 (300 Hz), 20 (1000 Hz), 26 (1300 Hz), and 36 (1800 Hz), respectively, if the most significant ones are indicated.

For the SPWM2 strategy (Fig. 16), the most significant electrical harmonics (currents) are shown in Fig. 13, which makes it possible shows that $fr = |\pm(a \cdot f_{sw} \pm b \cdot f) - f|$ equation is fulfilled. The electrical spectrum terms 15f, 25f, 33f, and 39f are located in the acoustic spectrum at frequencies of 800, 1300, 1700, and 2000 Hz. The remainder could not be identified due to the limitation of the analyzer on the first 50 electrical harmonics of the spectrum.

The SPWM3 strategy has electrical current harmonics that are significant for harmonics 5, 7, 15 and 49. Therefore, acoustic noise is expected at frequencies of 300 Hz, 400 Hz, 750 Hz, and 2450 Hz. For example, if a = 3 and b = 4, the frequency is 3·750+4·50 Hz = 2450 Hz. See Fig. 17.

The proposed strategy (HIPWM-FMTCt) has an electrical spectrum of line currents, as shown in Fig. 14. Fig. 18 illustrates the acoustic result with the HIPWM-FMTCt strategy for a control parameter value of K = 0.55 and 15 pulses per period. The harmonics 8f (400 Hz), 12f (600 Hz), 17f (850 Hz), 20f (1000 Hz), 25f, 26f (1250 Hz and 1300 Hz), and 41f (2050 Hz) were easily distinguished.

Table 3 illustrates the results obtained in the laboratory. The first two columns show the average SPL RMS of a 10 s measurement measured in the semi-anechoic chamber when the motor is excited by the different strategies at a line voltage of the fundamental term (50 Hz) of 220 V, and the value of the voltage THD obtained by the network analyzer. The third entry in the Table 3 represents the RMS value of the voltage obtained at the inverter output using the different control strategies when the DC voltage applied to the DC-link is 75V.

Table 3 has been configured in the noise level column from the mean value of 50 samples. The sonograms correspond to one of the closest samples for each technique. The results confirmed that the noise level emitted by the electric machine

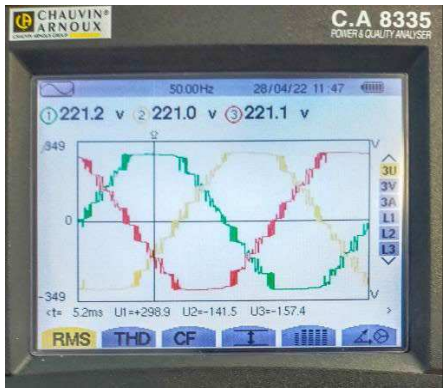

(a)

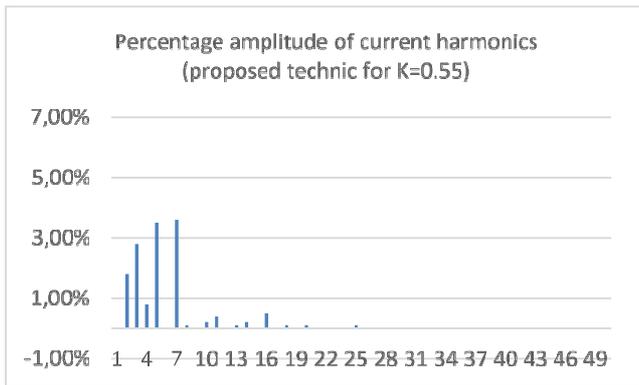

(b)

**FIGURE 14.** HIPWM–FMTCt with K = 0.55 and V = 220 v RMS of the fundamental term: a) Waveform at the output of the multilevel inverter for the HIPWM–FMTCt; b) Percentage frequency spectrum: current.

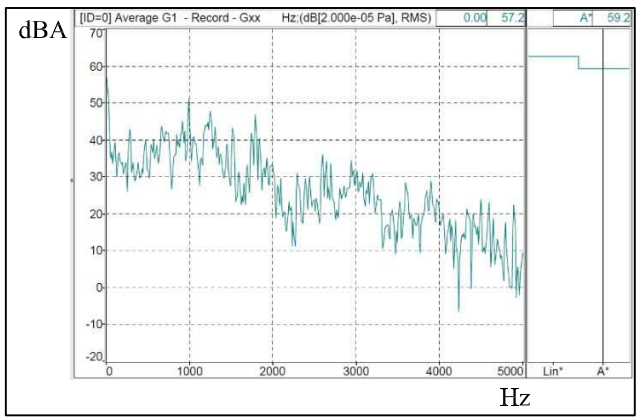

**FIGURE 15.** Acoustic response –dBA/Hz– for SPWM1.

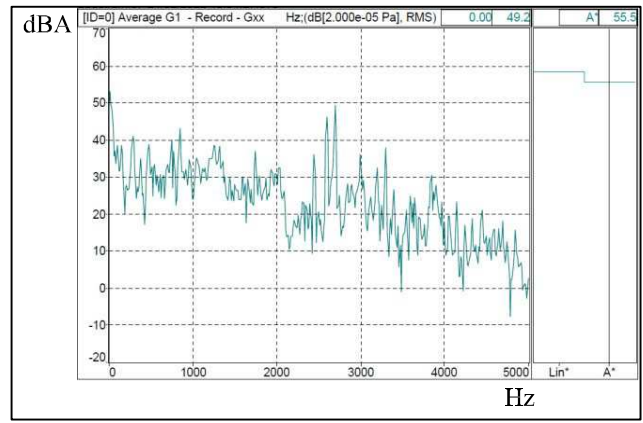

**FIGURE 16.** Acoustic noise emitted –dBA/Hz– with SPWM2 strategy.

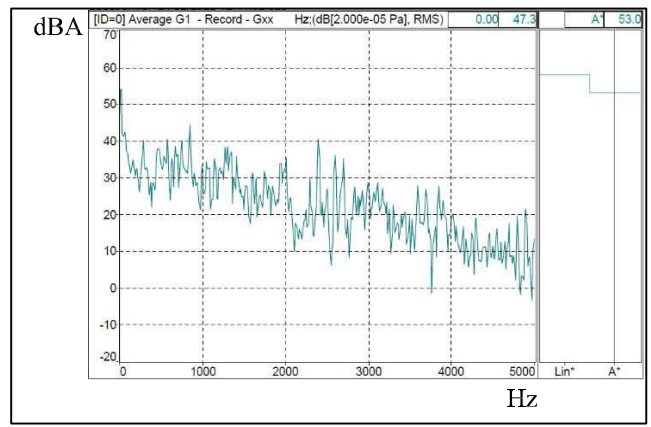

**FIGURE 17.** Acoustic spectrum emitted –dBA/Hz– with SPWM3 strategy.

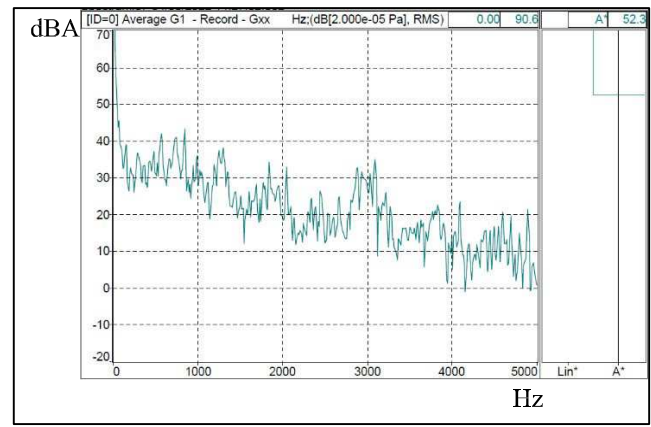

**FIGURE 18.** Acoustic spectrum –dBA/Hz– with HIPWM–FMTCt strategy for $K = 0.55, M = 15, A_M = 15\pi/0.45 = 33.3333\pi$.

for a reference value of $f_{SW} = 750$ Hz ($\overline{M} = 15$) was reduced over a wide range of K values (approximately K = 0.5). For small values of K, the truncated function of the instantaneous frequency of the carrier increases in width (fig. 3.b). Therefore, switching occurs outside the quasi-linear range of the modulating signal (figure 3.a). Therefore, the THD increases. When K values are close to the upper limit, the instantaneous frequency of the carrier wave increases, and therefore the frequency of the power IGBTs. When using real inverters and taking into account their dead time, the output waveform of the power inverter loses quality and thus the THD increases. As a conclusion, values close to K = 0.5 are the ones that improve the THD.

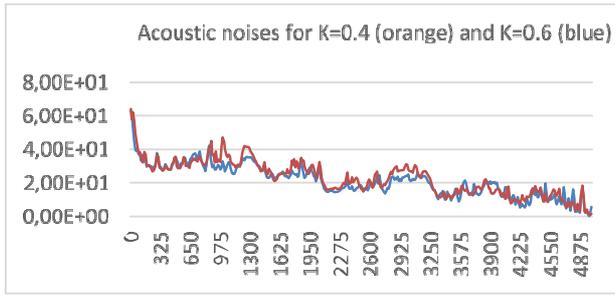

**FIGURE 19.** Comparison between noise harmonic levels for K = 0.4 and K = 0.6 ($M = 15$) with the HIPWM–FMTCt technique.

**TABLE 3.** Experimental results.

| Type of PWM/Power Supply | Level of noise dB(A) 220 V RMS fund. term | THD Voltage (%) 220 V RMS fund. term | $V_{RMS}(V)$ DC-link 75 V |
|---|---|---|---|
| Power supply | 49.6 | 0% | - |
| SPWM1 | 59.4 | 15.0% | 187 |
| SPWM2 | 55.7 | 4.5% | 190 |
| SPWM3 | 53.5 | 5.7% | 226 |
| HIPWM-FMTCt K=0.3 | 56.3 | 8.5% | 220 |
| HIPWM-FMTCt K=0.4 | 54.9 | 7.0% | 223 |
| HIPWM-FMTCt K=0.45 | 53.8 | 4.0% | 226 |
| HIPWM-FMTCt K=0.5 | 52.9 | 4.2% | 231 |
| HIPWM-FMTCt K=0.55 | 52.3 | 4.5% | 229 |
| HIPWM-FMTCt K=0.6 | 52.3 | 5.0% | 228 |
| HIPWM-FMTCt K=0.65 | 52.1 | 5.2% | 230 |
| HIPWM-FMTCt K=0.7 | 52.0 | 5.5% | 230 |
| HIPWM-FMTCt K=0.75 | 52.7 | 6.8% | 231 |
| HIPWM-FMTCt K=0.8 | 53.0 | 7.4% | 231 |

Finally, Fig. 19 graphically compares the acoustic results with the proposed technique for K = 0.4 and K = 0.6, showing that with the second value, the acoustic results are 2.6 dBA lower. Consequently, the value of K, which is a good compromise between the acoustic noise level and the THD parameter, can be identified for each engine.

## V. CONCLUSION

A PWM technique with the frequency modulation of the carrier wave synchronized with the modulating wave was simulated and implemented in a CHB multilevel inverter. Laboratory measurements were performed for both acoustic levels and electrical results.

The advantage of the proposed technique is that by using a K parameter, for the same number of switch pulses, the output waveform characteristics can be modified. Depending on the objective: minimizing noise, or THD or increasing the RMS value of the output voltage, there will be a more suitable K value. This can be seen in Table 3 of the paper where for a value of K = 0.7, the noise is improved. With K = 0.5, the THD is improved, and with higher values of K, the RMS value at the inverter output is increased. In this paper we have focused on choosing K values that reduce the harmonic MMFs of the motor and consequently the noise emitted without significantly affecting the rest of the electrical parameters. If the number of pulses is kept constant, the two variables K and $A_M$ can modify the electrical spectrum of voltage at the inverter output. It is possible to apply values of K that reduce the acoustic noise emitted by the motor at the same number of pulses per period by up to 1 dB(A) compared with other techniques in the technical literature. The acoustic level of the emitted noise could be reduced by avoiding mechanical resonances and spatial harmonics with a high winding factor in the tested motor. For the proposed technique with central K values, the THD value was lower than that obtained for the techniques with which it was compared.